\documentclass[draft]{agujournal2019}
\usepackage{url} 
\pdfoutput=1
\usepackage{amsmath}
\usepackage{soul}
\graphicspath{{./figs/}}
\makeatletter
\def\input@path{{./figs_new/}}
\makeatother
\newcommand{\ig}[2]{\includegraphics[width = #1]{#2}}

\draftfalse

\journalname{Geophysical Research Letters}

\begin{document}

\title{Soft X-ray imaging of Earth's dayside magnetosheath and cusps using hybrid simulations}

%
%




\authors{J. Ng \affil{1,2}, B. M. Walsh \affil{3}, L.-J. Chen\affil{2}, Y. Omelchenko \affil{4,5}}
\affiliation{1}{Department of Astronomy, University of Maryland, College Park, MD 20742, USA}
\affiliation{2}{NASA Goddard Space Flight Center, Greenbelt, MD 20771, USA}
\affiliation{3}{Boston University, Boston, MA 02215, USA }
\affiliation{4}{Trinum Research Inc., San Diego, CA 92126, USA}
\affiliation{5}{Space Science Institute, Boulder, CO 80301, USA}
\correspondingauthor{Jonathan Ng}{jonng@umd.edu}







\begin{keypoints}
\item Soft X-ray images are reconstructed using the results of hybrid simulations and virtual spacecraft based on the upcoming LEXI mission. 
\item The cusp, bow shock and magnetopause motion are captured by the X-ray emission. 
\item Kinetic structures such as foreshock cavitons can be seen in the X-ray intensity for short integration times.
\end{keypoints}

\begin{abstract}
Interactions between solar wind ions and neutral hydrogen atoms in Earth's exosphere can lead to the emission of soft X-rays. Upcoming missions such as SMILE and LEXI aim to use soft X-ray imaging to study the global structure of the magnetosphere.  Although the magnetosheath and dayside magnetopause can often be driven by kinetic physics, it has typically been omitted from fluid simulations used to predict X-ray emissions.
  We study the possible results of soft X-ray imaging using hybrid simulations under quasi-radial interplanetary magnetic fields, where ion-ion instabilities drive ultra-low frequency foreshock waves, leading to turbulence in the magnetosheath, affecting the dynamics of the cusp and magnetopause. We simulate soft X-ray emission to determine what may be seen by missions such as LEXI, and evaluate the possibility of identifying kinetic structures. While kinetic structures are visible in high-cadence imaging, current instruments may not have the time resolution to discern kinetic signals. 
\end{abstract}

\section*{Plain Language Summary}

When heavy solar wind ions and exospheric neutrals near the Earth interact with each other, X-rays can be emitted. Upcoming missions such as SMILE and LEXI will use these X-rays to image Earth's magnetosphere. Earlier simulations used to predict how these images appear have used fluid models which omit small-scale physics. In this work we use hybrid simulations, which include ion kinetic effects, to make predictions about X-ray images. The additional structures due to the kinetic physics can be seen when the images are taken at high cadence, but current imagers may not have the sensitivity to see these structures.

\section{Introduction}

Charge exchange interactions between high charge-state solar wind ions and neutrals in Earth's exosphere can lead to the emission of soft X-ray photons. For instance, the interaction
\begin{equation}
\text{O}^{+7} + \text{H} \rightarrow \text{O}^{6+*} + \text{H}^{+},
\end{equation}
in which an electron is transferred from an exospheric hydrogen atom to an oxygen ion, leaves the oxygen ion in an excited state. The decay of the electron to a lower energy state then leads to the emission of a soft X-ray photon. In the plasma environment around the Earth, soft X-rays are emitted due to the interaction of heavy solar wind ions and neutral hydrogen, as shown by various observational studies (e.g.~\cite{freyberg:1998, cravens:2001, carter:2008, carter:2011}).

The study of global magnetospheric dynamics using soft X-rays with wide field-of-view imaging has been suggested \cite{walsh:2016,collier:2012,branduardi:2012,kuntz:2015,sibeck:2018}. Spacecraft such as the Solar wind-Magnetosphere-Ionosphere Link Explorer (SMILE) \cite{branduardi:2018} and the Lunar Environment heliospheric X-ray Imager (LEXI) \cite{walsh:2020} will be launched in the near future. 
Potential results of soft X-ray imaging have been studied by using data from Magnetohydrodynamics (MHD) simulations \cite{kuntz:2015,sun:2015,walsh:2016,sun:2019,connor:2021,matsumoto:2022}. In these works it has been shown that it may be possible to observe physical phenomena such as Kelvin-Helmholtz instabilities \cite{sun:2015}, the magnetospheric cusps, and magnetopause motion and erosion due to reconnection \cite{walsh:2016,sun:2019,connor:2021,samsonov:2022,samsonov:2022b}. These results have shown how imaging can contribute to the understanding of global magnetospheric dynamics.

Due to the fluid nature of the MHD model, kinetic structures are not captured by these simulations. Hybrid simulations, with kinetic ions and fluid electrons, have been used to study the effects of ion kinetic physics on the global behaviour of the magnetosheath and magnetosphere. For example, ion-ion instabilities in the foreshock generate ultra-low frequency (ULF) waves, which may evolve into structures such as short large amplitude magnetic structures (SLAMS) \cite{schwartz:1992,chen:2021}, cavitons \cite{kajdivc:2011} and spontaneous hot flow anomalies  \cite{zhang:2013}. Other kinetic structures, such as high speed jets \cite{hietala:2012,raptis:2020}, are also present in these simulations \cite{omelchenko:2021jet,ng:2021}. These structures can cause large scale perturbations in the magnetosheath and at the magnetopause, sometimes even triggering reconnection \cite{lin:2005, omidi:2016,ng:2021,chen:2021jet}. Hybrid simulations have also been used to study the effects of foreshock waves on field-line resonances \cite{shi:2021}, and cusp energetic ions \cite{wang:2009}. 
It is important to understand which aspects of kinetic physics can be captured by X-ray imaging with present and future instruments.    

In this paper we discuss the results of reconstructing X-ray emission using global hybrid simulation data. In Section~\ref{sec:model} we discuss the hybrid simulations and the X-ray emission model used to calculate the X-ray emission intensities. In Section~\ref{sec:results} we discuss the images produced and the physical phenomena that can be seen, followed by a discussion of the limitations and a summary in Section~\ref{sec:summary}.

\section{Simulation and Soft X-ray model}
\subsection{Hybrid simulations}
\label{sec:model}

In this work we use data from hybrid particle-in-cell simulations to reconstruct soft X-ray emission intensity. Ions (in this case protons) are evolved kinetically as macroparticles while the electrons are treated as a charge-neutralising fluid. The governing equations of the HYPERS model and numerical details are available in \cite{omelchenko:2012, omelchenko:2021jet,omelchenko:2021}.

We discuss results using the same data from simulations in \citeA{ng:2022hybrid}. In both cases, the IMF is quasi-radial, with cone angle $170^\circ$, where the $B_x$ component of the IMF points towards the Earth. The clock angles are $0^\circ$ and $180^\circ$.  The simulations use a physical domain of 716$d_i\times$1334$d_i\times$1334$d_i$, where $d_i$ is the proton inertial length in the solar wind. The upstream ratio between thermal and magnetic pressure for both protons and electrons is $\beta =  8\pi n_0 T_0/B_0^2 = 0.5$, and the ratio of the speed of light to the upstream Alfv\'en speed is $c/v_A = 8000$. Here $n_0$, $B_0$ and $T_0$ are the solar wind values of density, magnetic field and temperature respectively. The Earth's magnetic dipole is tilted 11.5$^\circ$ sunward. The computational domain is covered by a 300$\times$500$\times$500 cell stretched mesh, with a uniform patch around the Earth being covered by 175$\times$300$\times$300 grid cells with a resolution of 1 cell per $d_i$. Outside the uniform patch, the mesh cells are stretched exponentially with a growth factor of 12. 

We use the Geocentric Solar Magnetospheric (GSM) coordinate system, where the $x$-axis points from the Earth towards the Sun, the $y$-axis points in the dawn-dusk direction and the $z$-axis completes the right-handed system. The initial Alfv\'en Mach number is 8 (i.~e.~$v_0/v_A = 8$). The dipole strength is scaled such that the nominal magnetopause standoff distance is approximately $100$ $d_i$ (using the solar wind value of $d_i\approx 100 \text{km}$). The inner boundary is a hemisphere of radius $50$ $d_i$ with absorbing boundary conditions for particles and perfectly conducting boundaries for fields. We note that the standoff distance of $100$ $d_i$ is smaller than the realistic standoff distance ($\sim 500$ $d_i$), but is much larger than the minimum distance needed to simulate an earthlike magnetosphere \cite{omidi:2004}. As was demonstrated \cite{toth:2017}, the global magnetospheric solution is not strongly sensitive to this scaling factor, with the ion inertial length dynamics of the numerical model occurring in a self-similar manner with respect to the original plasma system. In the simulation, the solar wind density is set to be $5$ cm$^{-3}$, and the magnetic field is $|B| \approx 3.8$ nT. The solar wind proton temperature is 3.6 eV. Simulation outputs are written every $3.125/\Omega_{ci} \approx 8.3\text{s}$ where $\Omega_{ci}$ is the ion cyclotron frequency calculated using the IMF. The conversion to  physical units is discussed in the Appendix.

Simulations with these parameters have been used in previous works \cite{omelchenko:2021jet,chen:2021jet,ng:2021,ng:2022hybrid} to study the effects of the quasi-radial IMF conditions on the magnetosheath and magnetopause. As this paper is focused on the results of soft X-ray imaging rather than an analysis of the plasma behaviour, we briefly summarize the results here. Under quasi-radial IMF, the conditions are favourable for the formation of foreshock kinetic waves and structures including high-speed jets -- regions of enhanced dynamic pressure in the magnetosheath. It should be noted that periods of quasi-radial (cone angle $< 30^\circ$) IMF, where the kinetic effects are strongest, occur $\sim 16\%$ of the time based on observations at 1 AU \cite{suvorova:2010}. In \citeA{omelchenko:2021jet}, jets in the simulations were identified and characterised and compared to observations. In \citeA{chen:2021jet} and \citeA{ng:2021}, the effects of foreshock turbulence and high-speed jets on the magnetopause were studied and shown to cause magnetopause indentations and trigger magnetopause reconnection. \citeA{ng:2022hybrid} studied the overall structure of the magnetosheath and the role of foreshock turbulence. For the slightly northward IMF case, it was shown that a dynamic plasma depletion layer developed, while in the slightly southward IMF case, the combination of foreshock turbulence and reconnection led to density enhancements in the cusp. 

\subsection{Soft X-ray model}

Following \citeA{kuntz:2015,connor:2021, cravens:2001}, the soft X-ray intensity is calculated using the line integral

\begin{equation}
  R_{xray} = \frac{\alpha}{4\pi}\int N_p N_N v_{\text{eff}}\, ds \left[\text{eVcm}^{-2}\text{s}^{-1}\text{sr}^{-1}\right],
  \label{eq:xray}
\end{equation}
where $v_{\text{eff}} = \sqrt{v_p^2 + v_{th}^2}$ is the effective velocity with $v_p$ and $v_{th}$ the bulk and thermal velocities respectively. $N_p$ is the ion density and $N_N$ is the neutral density. The integral is taken along the line-of-sight of the soft X-ray imaging device. We use $\alpha = 6\times 10^{-16}$ eVcm$^2$ \cite{connor:2021,cravens:2001}, and the neutral density is determined by
\begin{equation}
N_N = 25\left(\frac{10 R_E}{R}\right)^3\left[\text{cm}^{-3}\right],
\end{equation}
where $R$ is the distance from the centre of the Earth and $R_E$ is the Earth radius.

\section{Results}
\label{sec:results}

\begin{figure}
  \centering
  \ig{3.375in}{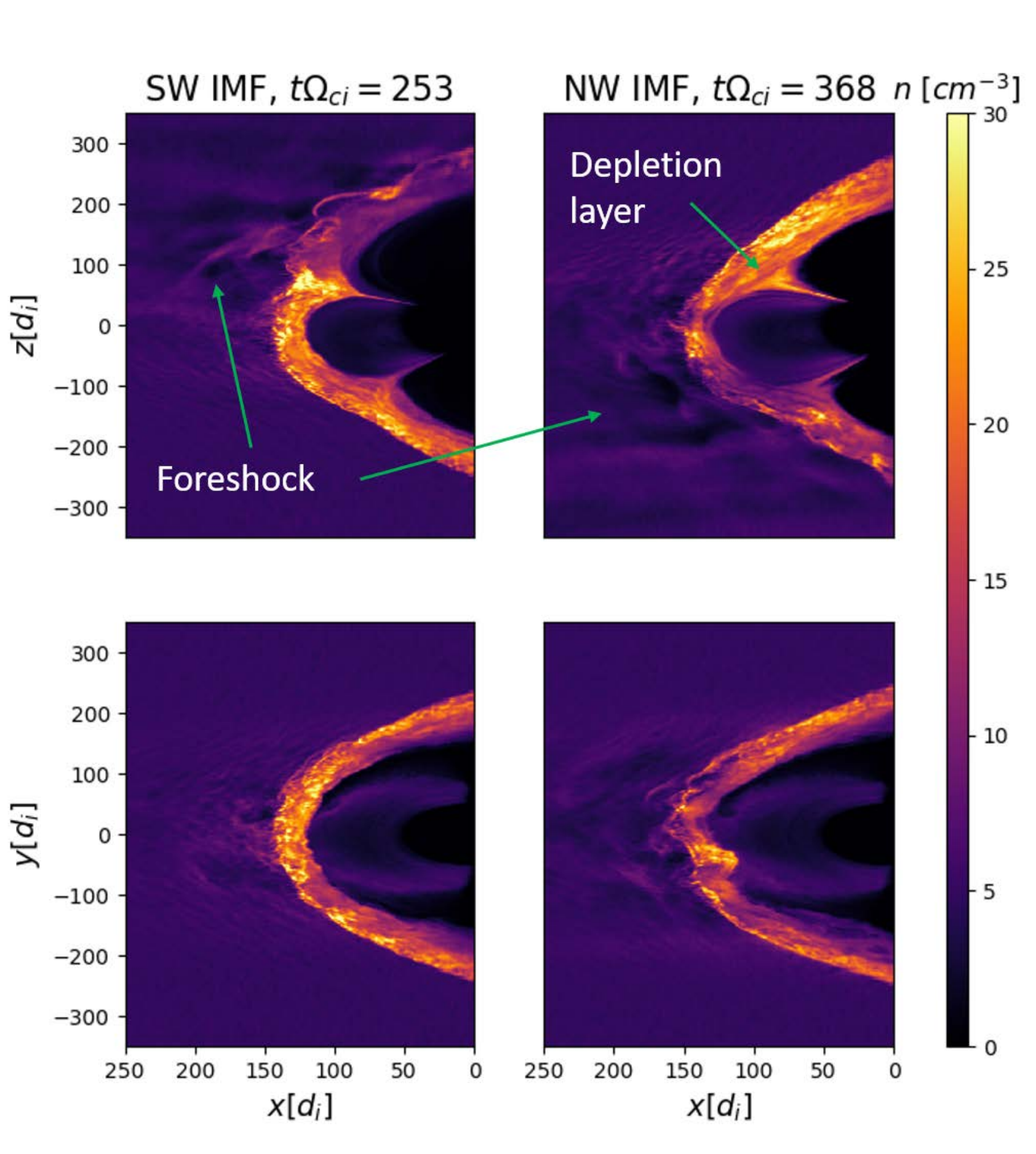}
  \caption{Overview of the quasi-radial IMF simulations with weak southward (left) and northward (right) components in GSM $x$-$z$ (top) and $x$-$y$ (bottom) planes using a subset of the domain. The aspect ratios of the axes are slightly distorted so that the entire dayside is shown. }
  \label{fig:overall}
\end{figure}

We first provide an overview of the simulation. Figure~\ref{fig:overall} shows the plasma density in the GSM $x$-$z$ and $x$-$y$ planes, with the southward IMF case on the left, and the northward IMF case on the right. The results shown in the figure are chosen to highlight specific features in the system. In both cases, there are strong foreshock density fluctuations due to the nonlinear evolution of ULF waves, and the magnetosheath is turbulent. The southward IMF case shows a density  enhancement outside the northern cusp associated a high-speed jet \cite{ng:2022hybrid}, while the northward IMF case shows a thin region of reduced density at the northern cusp boundary referred to as a plasma depletion layer (PDL) \cite{zwan:1976,lavraud:2005}. In the $x$-$y$ plane of the northward IMF simulation, a strong indentation of the magnetopause is seen in the $-100 < y < 0$ region, caused by the impact of a large high-speed jet approximately $10\Omega_{ci}^{-1}$ before. The time evolution of the systems can be seen as movies in \cite{ng:2022movie}.

\begin{figure}
  \centering
  \ig{4.375in}{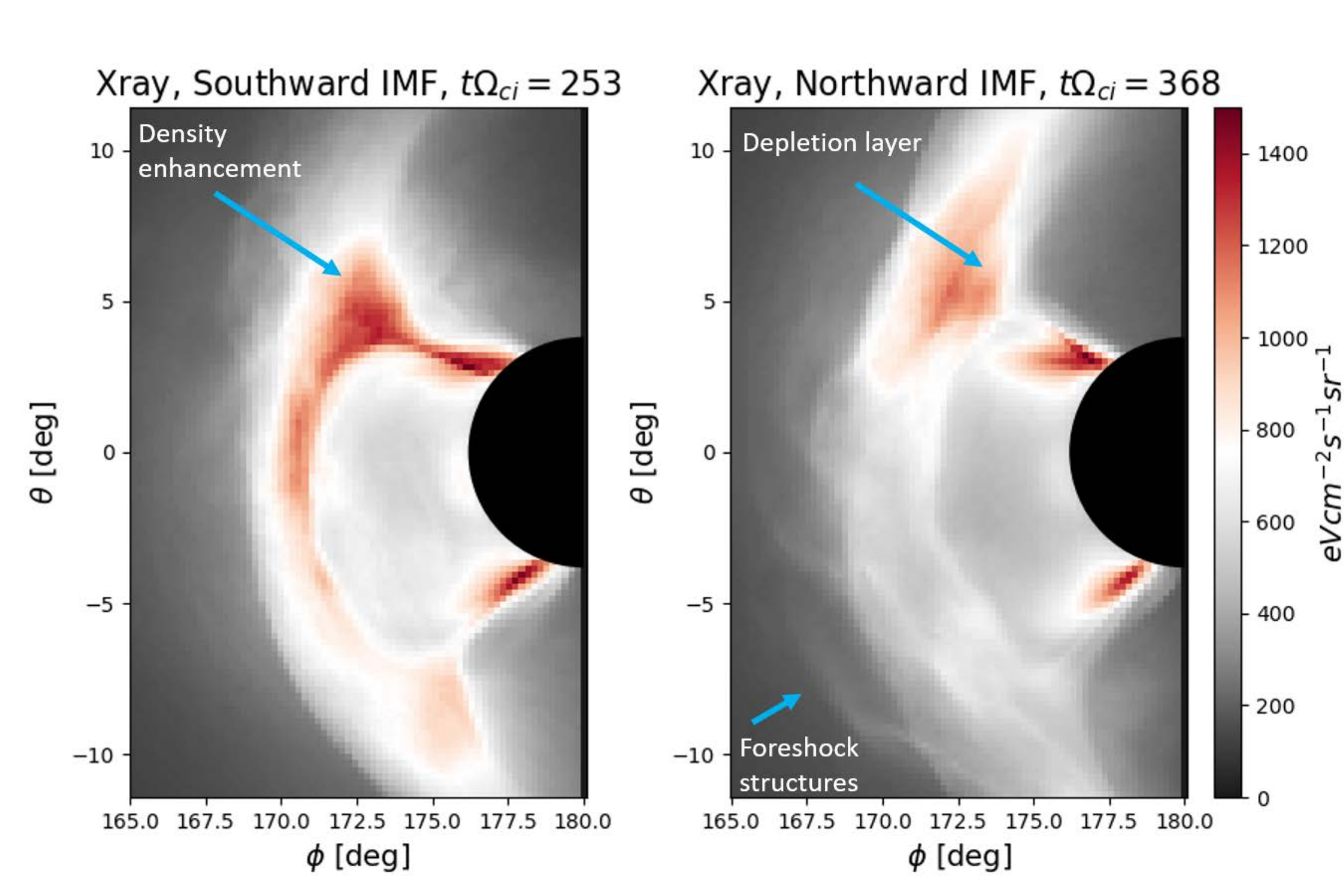}
  \caption{Soft X-ray images from the slightly southward and slightly northward IMF simulations using a detector positioned at GSM $y = 60 R_E$ with angular resolution $0.2^\circ \times 0.2^\circ$. }
  \label{fig:xray_overall}
\end{figure}

Using Equation~\eqref{eq:xray}, we calculate the soft X-ray intensities observed by a spacecraft located at $(x,y,z) = (0,60 R_E,0)$. The angular resolution of the images is $0.2^\circ \times 0.2^\circ$, corresponding to the resolution of the LEXI detector and the approximate orbit radius of the camera which is positioned on the moon. Note that the use of a stretched mesh reduces the accuracy of the integration away from the central domain, but the effect on the integral is small because of the $1/R^3$ scaling of the neutral density. The results are shown in Figure~\ref{fig:xray_overall}, where $\pi/2 - \theta$ is the polar angle measured from the positive $z$ axis, and $\phi$ is the azimuthal angle measured from the positive $y$ axis. The inner boundary is marked by the black area. In these results we have not performed masking of the magnetosphere as in \cite{samsonov:2022}, but the density in the magnetosphere and contribution to the X-ray intensity is sufficiently small that we can readily identify the magnetopause boundary in the results. 
In the southward IMF case, there is increased intensity in $\theta > 0$ region, corresponding to the enhancement seen in Figure~\ref{fig:overall}. While some of the increase in intensity can be attributed to the dipole tilt in the system (e.g.~\cite{walsh:2016}), the peak intensity in this region varies by up to a factor of two during the simulation, indicating the transient behaviour. The time evolution of this region close to the northern cusp can be seen as a movie in the Supporting Information. 

For the northward case shown in the right panel of Figure~\ref{fig:xray_overall}, there are clear physical features which are marked by arrows. Around $\theta = 5^\circ$, the intensity shows a minimum before the transition to the cusp close to $\phi = 173^\circ$, indicative of the PDL, while there are curved intensity enhancements in the foreshock region. These are related to the evolution of foreshock waves into structures known as cavitons or spontaneous hot flow anomalies \cite{lin:2005,blancocano:2009,zhang:2013}, characterised by high densities and magnetic fields at the rims and increased temperatures inside. One of the more prominent structures is found at $\theta = -7.5^\circ$, where there is a local maximum of intensity. Similar structures can be seen in the southward IMF case in the $\theta > 5^\circ$ region. Evidence of foreshock activity is visible in other regions just outside the magnetosheath where there is increased intensity compared to regions further away in the solar wind.

\begin{figure}
  \centering
  \ig{3.375in}{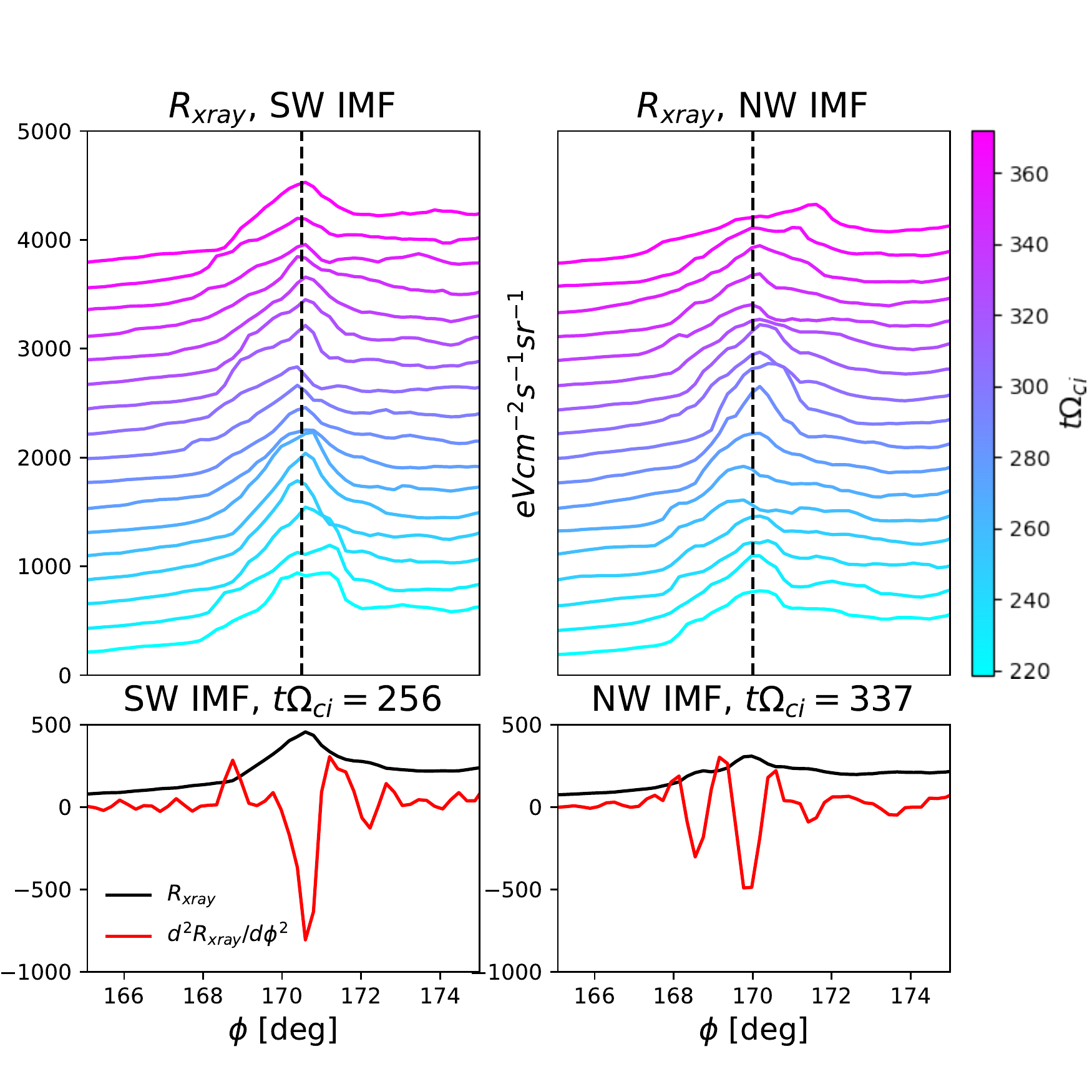}
  \caption{(Top) Stack plots showing the time evolution of the intensity at $\theta = 0^\circ$ in both simulations showing the motion of the magnetopause. (Bottom) X-ray intensities and $d^2R_{xray}/d\phi^2$ at selected times to show (left) how the bow shock and magnetopause locations can be determined, and how density inhomogeneities complicate the interpretation of the signal. A dashed vertical line is provided as a visual aid. The time interval between each trace is 9.375/$\Omega_{ci}$. }
  \label{fig:motion}
\end{figure}

The soft X-ray diagnostic can be used to study the motion of the magnetopause. As shown in \cite{collier:2018,sun:2020,connor:2021,samsonov:2022,wang:2022smile}, the peak intensity corresponds to the angle at which the line of sight is tangent to the magnetopause. As $\phi$ increases, the line of sight moves from the magnetopause to the magnetosphere, the intensity is reduced as the integral passes through a shorter region of the high density magnetosheath and parts of the low density magnetosphere. When $\phi$ decreases and the line of sight moves towards the bow shock, the intensity reduces as the path through the magnetosheath also shortens. Beyond point where the line of sight is tangent to the bow shock, there is also a change in the gradient as the line of sight no longer passes through the magnetosheath, leading to a reduced signal and less variation aside from foreshock activity. In addition to the maximum of intensity at the magnetosheath, there are changes in the gradient of the intensity at the transitions between the magnetosphere and magnetosheath, and the magnetosheath and the solar wind.  
This can then be used to track the motion of the magnetopause throughout the simulation. The location of the tangent to the bow shock is where the gradient of the intensity changes from the solar wind to the magnetosheath, and this is detected by finding the peak $d^2R_{\text{xray}}/d\phi^2$. The evolution of a cut along $\phi$ at $\theta = 0$ in both simulations is shown in Figure~\ref{fig:motion}, with a slight offset at each successive time for clarity. In both simulations, it can be seen that the peak moves, indicating magnetopause motion. This is caused by the changing plasma pressure in the turbulent magnetosheath. At certain times, the peaks broaden, likely due to magnetosheath inhomogeneities changing the density profile and hence the X-ray emission. At the latest time in the northward IMF simulation, the intensity peak is particularly far to the right due to a magnetopause indentation. There are also indications of density enhancements at the bow shock related to foreshock activity (e.g.~\cite{lin:2005,omidi:2016}) which can be seen as  ``knees'' in the intensity plots around $\phi = 168^\circ$.

Because of foreshock activity and density variations in the magnetosheath, the $d^2R_{\text{xray}}/d\phi^2$ signal is not as clear as it is in MHD simulations. Examples from different times used to illustrate the potential issues are shown in the lower panels of Figure~\ref{fig:motion}. The bottom left panel shows a straightforward trace, where the peaks in $d^2R_{\text{xray}}/d\phi^2$ at $\theta \approx 169^\circ$ and $170.5^\circ$ show clear indications of the bow shock and magnetopause. There are some also fluctuations in the foreshock region indicative of the wave activity there. Overall, this trace is quite similar to the examples shown in \cite{connor:2021}. A more complex example is shown in the bottom right. Here $d^2R_{\text{xray}}/d\phi^2$ shows multiple minima and maxima due to density fluctuations in the magnetosheath. While it is still possible to identify the bow shock and magnetopause positions using the maximum at $\phi \approx 168^\circ$ and the minimum at $\phi \approx 170^\circ$, the more complex trace indicates that the interpretation of the signal can be more difficult than what is expected from MHD studies because of the turbulence in the magnetosheath. 

\begin{figure}
  \centering
  \ig{3.375in}{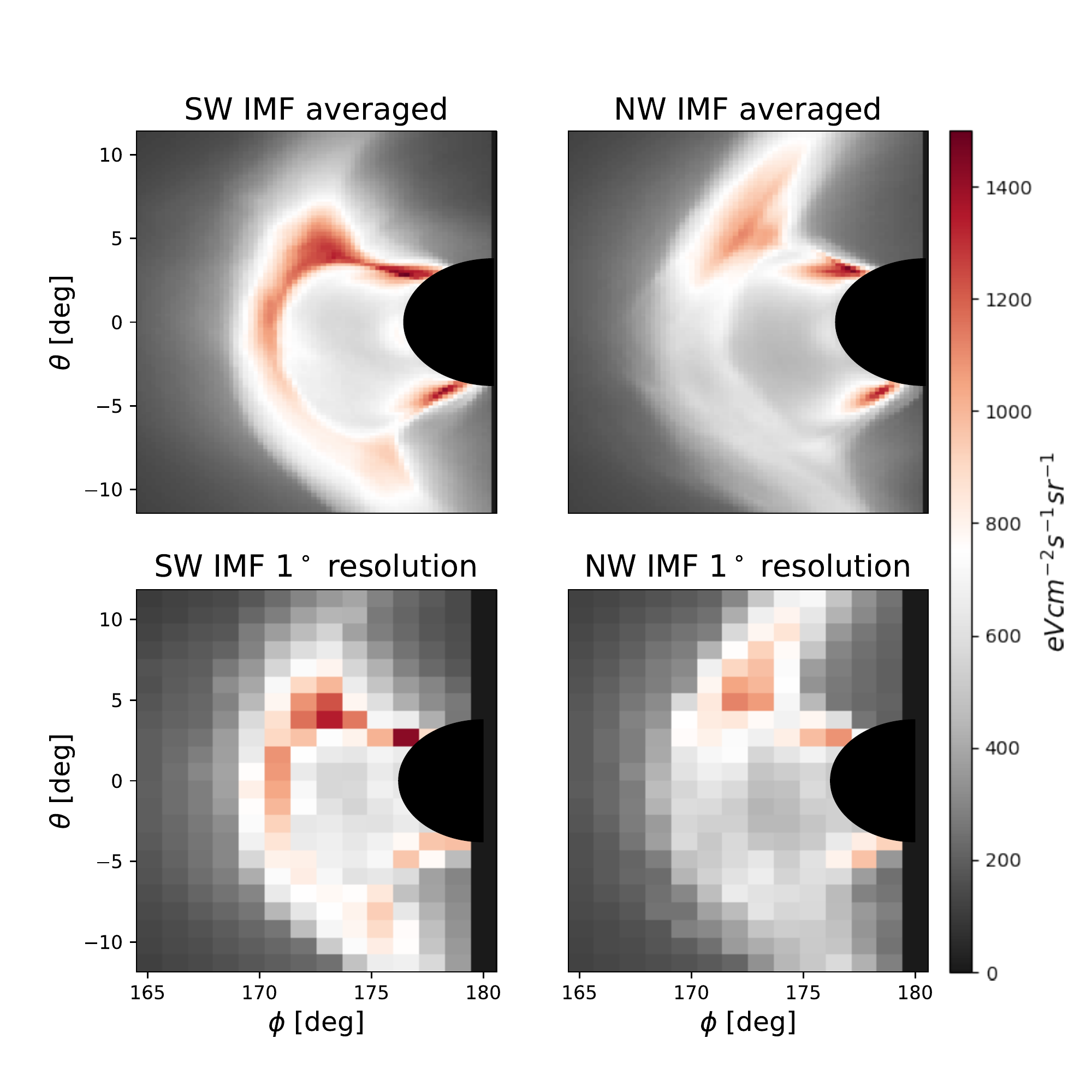}
  \caption{(Top) Time-averaged signals over 30 seconds in both simulations around the times used in Fig.~\ref{fig:xray_overall}. (Bottom) Lower-resolution images of the X-ray intensities at the times shown in Fig.~\ref{fig:xray_overall}. } 
  \label{fig:resolution}
\end{figure}

The question of whether the X-ray signal can be measured is an important one. So far, we have studied the ``ideal'' case, with the signal calculated using single outputs from the simulation. Fig.~\ref{fig:resolution} shows the results of time-averaging and using a lower angular resolution. Here the time-averaging is performed over approximately 30 seconds (corresponding to 4 simulation outputs around the times shown in Fig.~\ref{fig:xray_overall}). It can immediately be seen that the large-scale magnetosheath inhomogeneities are still present, though the fine structures in the foreshock are no-longer visible but are smeared out. Integrating over a more realistic longer period would thus further wash out the kinetic structures. Based on some experimentation with the simulation data, we find that the foreshock structures are visible when averaging over approximately 15 seconds is performed. The large-scale inhomogeneities are still visible with the time-averaging and with lower resolution. These plots do not include the contribution of the galactic background, that has been well-characterized by astrophysical observatories such as ROSAT \cite{trumper:1982, kuntz:2000}. While the magnetospheric signal varies on time-scales of minutes in response to the varying solar wind, the galactic contribution varies on much longer time scales, years or more, allowing much of it to be removed from the integrated signal \cite{sibeck:2018}. For a lunar-based observer looking away from the direction of the incident solar wind, the charged particle background is relatively low. The LEXI payload utilizes blocking filters and sweeping magnets to stop and repel charged particles. The modeled charged particle background for the payload is less than 1 count/s. A brief discussion of the results with added galactic background and Poisson noise is available in the Appendix. 

\section{Discussion and Summary}
\label{sec:summary}

We have used hybrid simulations of the dayside magnetosphere under quasi-radial IMF to calculate soft X-ray emission intensity. This technique will be used to image the magnetosphere by upcoming missions such as LEXI and SMILE, and in earlier work, MHD simulations have been used to show the possible physical phenomena that can be captured by the imaging method \cite{connor:2021,walsh:2016,kuntz:2015,sun:2015,sun:2019,matsumoto:2022,atz:2022}.

Under quasi-radial IMF, the physical conditions are favourable for foreshock turbulence to affect the subsolar regions of the magnetosheath and magnetopause. Our results show that some of these kinetic effects can be seen in the X-ray emission intensity, including transient density enhancements, magnetosheath indentations due to turbulence and  signatures of foreshock activity such as cavitons. Additionally, the structure of the intensity in the magnetosheath can be more complex than seen in MHD simulations because of kinetic turbulence. 

Whether the kinetic signatures can be detected by upcoming missions is less clear for a number of reasons. Because of the computational limitations of hybrid simulations, some scaling is required to convert the intensities of the reduced system to what would be expected from the Earth's magnetosphere, which is discussed in the Appendix. Imaging devices also have a finite integration time, meaning that transient structures may be washed out due to time-averaging. For instance, the foreshock structures in this work are visible when averaging over less than approximately 15 seconds, while the integration time for the SMILE mission is several minutes \cite{branduardi:2018}. More sensitive imagers in the future will likely be required to observe these structures. For these simulations, the solar wind density is relatively low, meaning that a longer integration time would be required to achieve a statistically significant signal to noise ratio, reducing the chances of seeing transient kinetic structures. Further kinetic studies with more favourable physical parameters such as higher densities and alternative IMF orientations will need to be performed to determine if the kinetic structures can be detected, and the conditions under which this will be possible.

To summarise, imaging using soft X-rays has been investigated using the results of hybrid simulations. The results show that kinetic structures can be seen in the X-ray emission signal, indicating the potential of this technique to observe kinetic physics. The ability of planned instruments to detect these kinetic structures will require further investigation. 


\acknowledgments

This work was supported by NSF grant AGS2010231, NASA grants 80NSSC21K1046, 80NSSC21K1483, 80NSSC23K0330.

\section*{Open Research}
Data used in this paper are available at \cite{ng:2023_xraydata}. 

\appendix

\section{Scaling of units to realistic system size}

As mentioned in Section~\ref{sec:model}, the hybrid simulation uses a reduced length scale due to computational limitations. This is typical in hybrid simulations, where a compromise between physical accuracy and computational cost has to be made by scaling the $\omega_{pi}/\Omega_{ci}$ ratio (ion plasma frequency to cyclotron frequency in the solar wind), the solar wind density or size of the Earth (both effectively control $d_i/R_E$) or under-resolving the simulation. In this work, we have chosen the solar wind density $n = 5$ cm$^{-3}$ to keep the ion cyclotron frequency reasonable, which effectively reduces the size of the Earth, though the system size is sufficient for the development of an Earth-like magnetosphere  \cite{omidi:2004,toth:2017}.

Because of this, the X-ray intensities are lower than in comparable MHD simulations \cite{connor:2021, walsh:2016, sun:2015}. From Equation~\eqref{eq:xray}, which we reproduce here for convenience, 
\begin{equation}
  R_{xray} = \frac{\alpha}{4\pi}\int N_p N_N v_{\text{eff}}\, ds \left[\text{eVcm}^{-2}\text{s}^{-1}\text{sr}^{-1}\right],
\end{equation}
the only physical quantity which is reduced is the length scale, or the $ds$ integral, which is reduced by an approximate factor of 5. One may perform a na\"ive scaling and multiply the intensities by this factor to obtain the expected intensities. If we compare the values from the figures in this paper to \cite{connor:2021}, we note that their magnetosheath signal is $\sim 15 keV cm^{-2} s^{-1} sr^{-1}$, compared to $\sim$4--7 $keV cm^{-2} s^{-1} sr^{-1}$ in this paper (Figures~\ref{fig:xray_overall} and~\ref{fig:resolution}).  However, noting that their solar wind density is 10 cm$^{-3}$, twice ours, we find that the X-ray intensities are overall consistent.

\begin{figure}
  \centering
  \ig{3.375in}{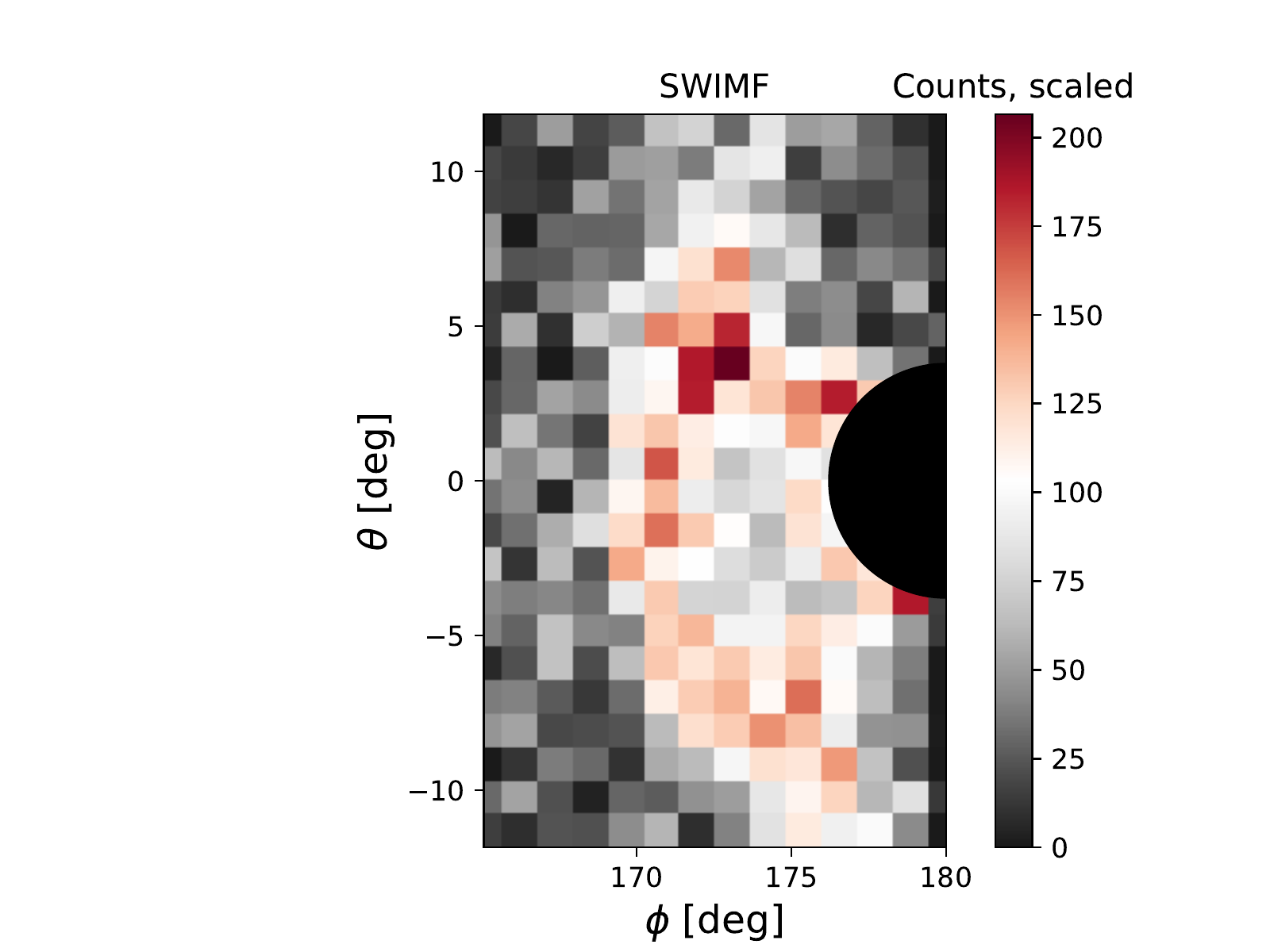}
  \caption{Possible detector output using LEXI instrument parameters and a $1^\circ \times 1^\circ$ resolution after integrating over 1200 seconds, after the addition of noise and background. }
  \label{fig:instrument}
\end{figure}

For comparing hybrid simulations to expected measurement results, it is necessary to use the scaled values to obtain reasonable predictions. An illustration of possible measurements from the southward IMF simulation is shown in Figure~\ref{fig:instrument}. To produce these measurements, we have used the calibration for LEXI to convert to X-ray counts, then added the galactic background and poisson noise, similar to \cite{walsh:2016}, and integrated over approximately 1200 seconds because the solar wind density is low (due to the limited duration of the available simulation data in quasi-steady state, results are averaged over 300 seconds, then scaled by a factor of 4). For this paper, the low density, which leads to a weaker X-ray signal, was chosen as the physics in the simulations with these parameters has already been analysed. Further work with a higher initial solar wind density will be necessary.

\bibliography{reconnectionbib.bib}

\end{document}